\documentstyle[prl,twocolumn,aps,epsf,epsfig]{revtex}
\begin{document}
\draft
%\documentstyle[prl,aps,preprint,tighten]{revtex}
%\tightenlines
%\documentstyle[prl,twocolumn,aps]{revtex}
%\topmargin -1.0cm
%\begin{document}

\title{Magnetic Domain Walls in Double Exchange Materials}

\author{Luis Brey}
\address{Instituto de Ciencia de Materiales de Madrid, 
CSIC\\
28049 Cantoblanco, Madrid, Spain.}

\date{\today}

\maketitle

\begin{abstract}
We study magnetic  domain walls in double exchange materials.
The domain wall width is proportional to the square root of the stiffness. 
In the double exchange model the  stiffness
has two terms: the kinetic energy and the Hartree term. The kinetic energy term 
comes
from the decrease of the tunneling amplitude in the domain wall region.
The Hartree term appears only in double exchange materials and it
comes from the connection between band-width and magnetization.
We  also calculate the low-field magnetoresistance associated
with the existence of magnetic domains. We find  a magnetoresistance of $1-2 \%$. The magnetoresistance can be considerably larger in magnetically constrained nanocontacts.

\end{abstract}
PACS number 71.10.-w, 75.10.-b, 72.10
%\tableofcontents

Mixed valence compounds of the form  
$La  _{1-x} A  _{x} Mn  O _{3} $
($A$ being  Ca, Sr or Ba) have recently been shown to present
extremely large (colossal) magnetoresistance.\cite{prim,rvw}
In these materials strong Hund's interaction between the charge carriers
and the manganese ions leads to a strong coupling between the electrical
resistivity and the magnetic state.
For $0.1\leq x \leq 0.5$ and low temperatures, 
the system is metallic and presents ferromagnetic order. 
As the temperature 
increases the system becomes insulator and paramagnetic. 
The magnetic transition occurs at a $x$-dependent critical temperature
$T_c \sim 300K$.  
Colossal magnetoresistance occurs near $T_c$ and in presence of magnetic fields
of several Teslas.

In the 
$La  _{1-x} A  _{x} Mn  O _{3} $
compounds,
the electronically active orbitals are  the
$Mn$ $d$ orbitals, and the mean $d$ occupancy is $4-x$. 
A  strong ferromagnetic 
Hund's rule coupling aligns all electron spins in the $Mn$ $d$ orbitals.
The $Mn$ ions form 
a simple  cubic lattice of lattice parameter $a$. The cubic crystal symmetry splits 
the $d$ orbitals into a $t_{2g}$ triplet
and an $e_g$ doublet. 
Three electrons fill up
the $t_{2g}$  levels forming a core spin  of magnitude $S=3/2$
and the rest of the electrons, $1-x$ per $Mn$, go to the $e_g$ orbitals.

Ferromagnetism in these materials is explained by the
Double Exchange (DE) mechanism, in 
which the electrons
get mobility between the $Mn$ ions using the oxygen
as an intermediate.\cite{zener,anderson,degennes} 
This conduction process is proportional to the electron transfer integral
and due to the strong ferromagnetic Hund's rule coupling 
it is maximum when the two cores spins involved in the process 
are parallel and it is zero
when they are antiparallel.
Because
the alignment of spins favors electronic motion, the
ferromagnetic ground state maximizes the electron kinetic energy. 
When the temperature increases, the DE model undergoes a phase
transition towards a paramagnetic state. In this phase the core spins are
randomly oriented and 
the electron kinetic energy is  minimized. 
In the paramagnetic phase these materials behave as electrical 
insulators.\cite{note1}

Large low-field magnetoresistance has been observed  in ferromagnetic 
$La   _{1-x} A  _{x} Mn  O _{3} $
compounds with different structural discontinuities.\cite{grain1,films1,tri1}
These effects are associated with a lack
of oxygen at the interface which produces an antiferromagnetic ordering
at the interface and breaks the DE mechanism.\cite{calderon1}

Below $T_C$ these materials
may contain magnetic domains separated by domain walls (DW's).
Domain walls produce a resistance to the electrical current, 
and 
for understanding  low-field magnetoresistive effects in
DE metals
it is important to know the width and the resistance 
of  DW's.
Domain wall magnetoresistance effects have been recently observed in itinerant
ferromagnet systems such as $Co$, $Ni$ and $Fe$\cite{iti,nico} and also
in colossal magnetoresistance
perovsquites.\cite{mathur}
In reference [13] it is obtained that in 
$La   _{1-x} Ca  _{x} Mn  O _{3}$ the resistance of a domain wall is
$8\times 10 ^{-14} \Omega \, m ^2$, a quantity that the authors 
argue is $4$ orders of magnitude larger
that one might expect based in DE models.
Recent theoretical works have studied the ballistic and diffusive transport
through domain walls in itinerant ferromagnets,\cite{itith} 
however for DE systems only the transmission in one-dimensional models 
have been studied.\cite{yamanaka}.

In this paper we study 
magnetic domain walls  in DE systems.
In the DW the direction of the $Mn$ spin changes from
$0$ to $\pi$ over a region of width $L_W$. In this region the $Mn$ core spins
are misaligned  and the tunneling amplitude between $Mn$ ions along
the DW is reduced. This loss of kinetic energy in the DE
model is the equivalent to the loss of exchange energy in the Heisenberg model.
The chemical potential in the system is fixed by the magnetic domains
which have all the same hole concentration $x$.
The reduction of the bandwidth in the DW region with respect
of the surround magnetic domains produces a change in the 
density of electrons in the DW region. This effect  cost a lot 
of Hartree energy and the system  prefers to create dipoles at 
the edge of  the DW's. In this way   the  
$Mn$ ion levels  change and the local charge is not modified
in  the DW.
The shift of the energy level of the $Mn$ ions modifies  the
cost  of creating a DW and therefore its  
width.
This effect is new and  occurs due to the DE mechanism.
In the first part of the paper we characterized this effect and study how 
it affects the width of the DW. 
In the second part of the paper we study the ballistic transport through
DW's in DE systems.

{\it Microscopic Hamiltonian.}
We are interested in a hole concentration in the range
$0.1  \leq x \leq 0.4 $. For  this doping range the $e_g$ orbitals 
are degenerated and for simplicity we consider that the Hamiltonian
is degenerated in the $e_g$ orbital index.
Also in our model the $Mn$  spins are treated as classical. 
For temperatures below $T_c$ and in the limit of infinite Hund's coupling, 
the electronic properties of the
$Mn$ oxides are described by the nearest neighbor tight binding
Hamiltonian,
\begin{equation}
\widehat H _{DE} = - \sum _{i,j, \alpha } 
t _{i,j} \widehat C ^ {+} _{i, \alpha} \widehat C _ {j , \alpha} 
\, \, \, \, \, ,
\end{equation}
Here $\widehat C^ {+} _{i,  \alpha} $ creates an electron at site $i$, 
in the orbital $\alpha$ and  
with spin parallel 
to the core spin at site $i$, 
and the hopping amplitude is given by\cite{muller} 
\begin{equation}
t_{i,j}= t \left ( \cos {{\theta _ i } \over 2}
\cos {{\theta _ j } \over 2}
+\sin {{\theta _ i } \over 2}
\sin {{\theta _ j } \over 2}
e ^ { i ( \phi _i - \phi _j )} \right ) \, \, \, ,
\label{tcomplex}
\end{equation}
where $\theta _ i$ and $ \phi _i$ are the angles which characterize
the orientation of the core spin at site $i$.

{\it Long-wavelength functional.}
In a perfect system all core spins are parallel and the electron kinetic 
energy gets its maximum value.
Now consider a modulation of the core spin direction  characterized by a vector
${\bf M} _i = {\bf S} _i / S$.
In the long-wavelength limit the modulation can be described by a 
continuum unitary
vector field ${\bf M}( {\bf r})$.
This modulation produces a decrease of the value of the hopping amplitude,
and therefore a  kinetic energy loss. 
For smooth modulations the local loss
of kinetic energy is proportional to 
$\left ( {\bf {\nabla}} {\bf M} \right ) ^ 2$
%the square of the divergence of the
%unitary vector field ${\bf M}$, 
and the difference in kinetic energy (KE)  between the 
uniform  system and the modulated  system 
is given by
\begin{equation}
\Delta E ^{KE} = { { \rho ^{KE} } \over 2} \int d ^3 {\bf r}
\left ( {\bf {\nabla}} {\bf M} \right ) ^ 2 \, \, \, \, .
\end{equation}
where $\rho ^{KE}$ is the  KE stiffness of the system.
This term is the equivalent to the term describing the exchange energy loss
in the Heisenberg model. 
Note, however that in the DE system $\rho ^{KE}$ is associated 
with  the loss of electron kinetic energy  due to the 
spatial variations of ${\bf M}$.
By diagonalizing the Hamiltonian Eq.1 with different boundary conditions we
have calculated the value of $\rho ^{KE}$ for different hole  concentration 
in the system.
In Fig.1 we plot $\rho ^{KE}$ as a function of  $x$.
Note that we have a degeneration 2 associated with the
$e_g$ orbitals and we measure the hole concentration with respect half-filling.
When
the electron concentration is zero ($x=1$)
the band is completely empty there is not 
KE in the system and  $\rho^{KE} =0$. 
When the band is half filled ($x=0$) 
the  KE is maximum and $\rho^{KE}$ gets its maximum value.

Besides the KE, there is also a Hartree (H) term which
contributes to the long-wavelength functional.
The electron density, $n$ ,in different regions of the sample is fixed by the 
chemical potential $\mu$ which  is obtained from the value of $n$ 
in the regions of constant magnetization (magnetic domains).
The tunneling amplitude only depends on the
relative orientation of the $Mn$   spins and therefore
$n$ and $\mu$ get the same values  within all 
the magnetic domains. 
These regions represent basically all the system and they can be  
considered as {\it reservoirs}  for the electrons.
The total charge in the system should be zero and 
a background of positive charge equal to $1-x$ 
is assumed to exist in the sample.

In regions with modulated magnetization, ${\bf \nabla}   {\bf M} \neq 0$,
the tunneling amplitude is reduced and  the bandwidth is narrowed.
Since $\mu$ is fixed by the {\it reservoirs}, the decrease of the 
bandwidth would produce a change in $n$ in these 
regions.
This breaks local charge neutrality and it would cost a lot of Hartree
energy. The system prefers to create dipoles in order to shift the $Mn$ ion
energy levels in such a way that local charge neutrality is recovered.
The energy shift is negative (positive) for electron concentration 
smaller (bigger) than half-filling.
For smooth modulation of the magnetization, we have calculated from the
Hamiltonian Eq.(1) the energy shift required for keeping local
charge neutrality. We find that the shift is proportional to
$\left ( {\bf \nabla}   {\bf M}  \right ) ^ 2$
and therefore the  local charge neutrality constrain gives a 
contribution to the 
energy of the system of the form,
\begin{equation}
\Delta E ^{H} = \beta  \int d ^3 {\bf r}
\left ( {\bf {\nabla}} {\bf M} \right ) ^ 2 \, \, \, \, .
\end{equation}

In Fig.1 we plot the value of $\beta$ as a function of $x$, 
as obtained by solving
the Hamiltonian Eq.(1) with different boundary conditions.
In principle we should add also a term describing the energy cost of
creating the dipoles responsible for the energy shifts, however we find 
that this energy is much smaller than $\Delta E ^{H}$.

Finally we need to know the energy term corresponding to the constrain 
which creates the magnetic domains. In general this term has the form,
\begin{equation}
\Delta E ^{C} =  \int d ^3 {\bf r} \, f(\theta)
\, \, \, \, .
\end{equation}
For uniaxial crystals $f(\theta) \!= \! K \sin ^ 2 (\theta) $ being 
$K$ the anisotropy constant. For domain
walls created by magnetic constrictions\cite{nico,mathur}
$f(\theta) \! = \! g \mu _B H S / a ^3 \, \cos{ \theta } \, \mbox{sgn} (x)$, being
$H$ the magnetic field.

Adding the different contributions we get the functional 
\begin{equation}
F=  
{ { \rho  } \over 2} \int d ^3 {\bf r}
\left ( {\bf {\nabla}} {\bf M} \right ) ^ 2 
+ \int d ^3 {\bf r} f(\theta)
\, \, \, \, .
\end{equation}
being $\rho = \rho ^{KE} + 2 \beta$.

{\it Domain wall width.}
A DW along the $\hat x$ direction has  the general form
${\bf M} = (0,\sin{\theta}, \cos{\theta})$  and 
$\cos {\theta (x)} $  is obtained by minimizing 
$F$ with the boundary conditions,
${\bf M} \rightarrow \pm  {\hat z}$ at $x \rightarrow \pm \infty$. 
For  uniaxial crystals
the optimum form of the DW is
\begin{equation}
\cos {\theta} = - \tanh { {{4 x}  \over {L_W}}}  \, \, \, \, 
\end{equation}
with $L_W = \sqrt{ \rho / 2K} $. For a magnetic constrain the form of the
domain wall is given by  
\begin{equation}
\cos {\theta} \! = \!
\left \{ 1 - 2 \cosh ^ {-2} \left [ 
\left| { { 4 x} \over {L_W}}  \right|  \!  +  \! 
\ln ( \sqrt {2} \! + \! 1 )  \right ] \right \}
\mbox{sgn} (x)
\, \,  \end{equation}
and $L_W= 4 \sqrt{ \rho  { { a ^3} \over { g \mu _B H S}} }$.
In both cases $L_W$ represents the width of the DW, and
it is proportional to the square root of the total stiffness 
$\rho$. The effect of the Hartree term to the DW width
depends on the electron concentration: for electron concentrations
smaller than half filling  $\beta $ is negative and the Hartree term
prefers thin DW's. On the contrary 
for electron concentrations bigger than half filling  $\beta $ is positive
and the Hartree term likes wide DW's.

For a given value of $L_W$ equations (7) and (8) 
have essentially the same form and
for simplicity we use expression (7) for describing a domain wall. 

{\it Transport through a domain wall}. Now we  calculate the 
ballistic conductance, $G$, associated with a DW.
From the difference between the  conductances  of a perfect system and 
a system with a DW
we can evaluate the low field magnetoresistance associated with the alignment 
of the magnetic domains.
A DW modulates the magnetization only along $\hat x$ direction and the
Hamiltonian  
is invariant in the  $\hat y - \hat z$ plane. 
Therefore the conductance of the system can be written as
\begin{equation}
G( \mu ) = \int d E ^{\prime} \, g _{1D} ( E ^{\prime} ) \, 
n^{2D} (\mu -E ^ {\prime} ) \, \, \, ,
\end{equation}
here $n^{2D} (E)$ is the two-dimensional density of states per unit area at energy
$E$,
% which for a square lattice has the form,
%\begin{equation}
%n^{2D} (E)= { 2 \over {4 \pi ^ 2 t a ^ 2}} 
%\Theta ( 4 t - |E|) K ( \sqrt { 1- { E  \over { 16 t ^ 2}} } )
%\end{equation}
%being $K$ the complete elliptic integral of the first kind.
and  $g _{1D} (E)$ is the conductance of a one dimensional 
system at energy $E$.
%The DW is contained in the one dimensional system.
Expression (9) can be interpreted as the sum of the conductance of all
the one dimensional channels. 
%being $n^{2D} ( E)$ the number of channels
%with energy $E$.
The one dimensional conductance between the sites $1$ and $N$ 
is written as,
\begin{equation}
g_{1D} ( \mu) = 2 { { e ^2} \over {h}} 4  \pi ^ 2 t ^ 4 D ^ 2 (\mu) 
|G _{N,1} (\mu)| ^ 2 \, \, \,
\end{equation}
being $D(\mu) = \sqrt { 4 t ^2 - \mu } /(2 \pi t ^2)$ the local
density of states at an edge of the isolated lead, and 
$G _{N,1} (\mu)$ the Green function connecting the sites $1$ and $N$
of an infinite 
one dimensional system.\cite{oguri,calderon2}
The factor 2 in Eq.(10) corresponds to the $e_g$ degeneracy.
The DW is fully contained between the sites $1$ and $N$.

In order to calculate the effect of the DW on the transport properties 
it is necessary to know the effect that the  magnetization,
${\bf M} = (0,\sin{\theta}, \cos{\theta})$ has on the electron Hamiltonian,
There are two effects: the modulation of the hopping amplitude along the 
$\hat x$ direction 
%$t \rightarrow t \cos { {{ \theta _i -\theta _j} \over 2}}$
and the shift of the $Mn$ ion levels.
Both effects are  obtained from the form of the DW, Eq.(7).

From Eq.(9) we  evaluate 
the magnetoresistance,
\begin{equation}
MR= {{ G_0 - G _{DW}} \over {G_0}} \, \, \, \, .
\end{equation}
Here $G_{DW}$ and $G_0$ are respectively the conductance in presence and in absence of the DW.
$MR$ represents  the low-field ballistic magnetoresistance associated 
with the presence of magnetic domains.
In Fig.2 we plot, for different hole concentrations, $MR$ as a function 
of $L_W$.
For small values of $L_W$, $MR$ gets rather large values ($> 10\%$). For
$L_W \geq 20 a$, the magnetoresistance is always smaller than $1 \%$.
In the inset of Fig.2 we plot the the one-dimensional conductance, $g_{1D}$ as
a function of the energy for a DW with $L_W$=$5a$. The asymmetry with respect 
to zero energy, is due to the Hartree energy  level shift.
We see that $g_{1D}$ is suppressed mainly at the band edges.This is 
the reason why the $MR$ is bigger for small concentration of electrons, where
the Fermi energy is close to the band edge.

{\it Discussion.} We now discuss the application of our results to manganese oxides.
First we want to know the width of the DW produced by crystal anisotropy. The width
is determined by the stiffness,  $\rho$, and by the anisotropy constant $K$.
$K$ can be obtained experimentally
from neutron scattering\cite{hirota,hwang,perring}, by 
microwave absorption\cite{sanchez} and by studies of the resistance
saturation with magnetic fields.\cite{films1}
From these experiments we  estimate an anisotropy constant in the range 
$2 \times 10 ^5 - 5 \times 10 ^6 J m ^{-3}$,
for $x=0.3$ this implies  a DW width in the range
$10a \leq L_W \leq 30 a$ and  a ballistic
MR of $1-2 \%$.

In the case of  DW's created by magnetic constrictions\cite{mathur}, $L_W$
depends on the applied magnetic fields. For $x=0.3$, we obtain 
$L_W \simeq 26 a / \sqrt {H}$ being $H$ the external magnetic field in Teslas.
This corresponds to a rather large DW. Therefore we expect that the DW width 
should be determined 
by the crystal anisotropy and $ \sim 1-2 \%$ magnetoresistances are expected. 
This value is similar to the obtained in reference [13], assuming 
$10a \leq L_W \leq 30a$.
However we do not know what is the contribution of diffusive processes
to $MR$. Within the simplest Born approximation, diffusive effects are
described by the reduction of the bandwidth, and for the values of
$L_W$ considered  this produces also a $1\%$  magnetoresistance. 

From the inset of figure 2, we can say that the magnetoresistive
effects should be much more important in geometrically constrained
domain walls. Recently magnetoresistances of 200$\%$ have been observed
in nanocontacts of itinerant ferromagnets.\cite{nico,bruno}. We expect similar
effects to occur in DE materials confined to reduced dimensions.\cite{ott}

{\it Summary.} We have calculated the width of domain walls in double exchange materials. 
The width is proportional to the square root of the stiffness. The stiffness
has two terms; the kinetic energy and the Hartree term. The kinetic energy
is the equivalent to the exchange energy in the Heisenberg model and it comes
from the decrease of the tunneling amplitude in the domain wall region. 
The Hartree term appears only in double exchange materials and it
comes from the connection between band-width and magnetization.
We have also calculated the low-field magnetoresistance associated
with the existence of magnetic domains. We have found a magnetoresistance of  $1-2 \%$. The magnetoresistance can be considerably larger in magnetically constrained nanocontacts.

We thank G.Platero, M.J.Calder\'on, J.A.Verg\'es and F.Guinea for useful discussions.
This work was supported by the CICyT of Spain under Contract No. PB96-0085
and by  the
Fundaci\'on Ram\'on Areces.

\vspace{-7cm}
\begin{figure}
\epsfig{figure=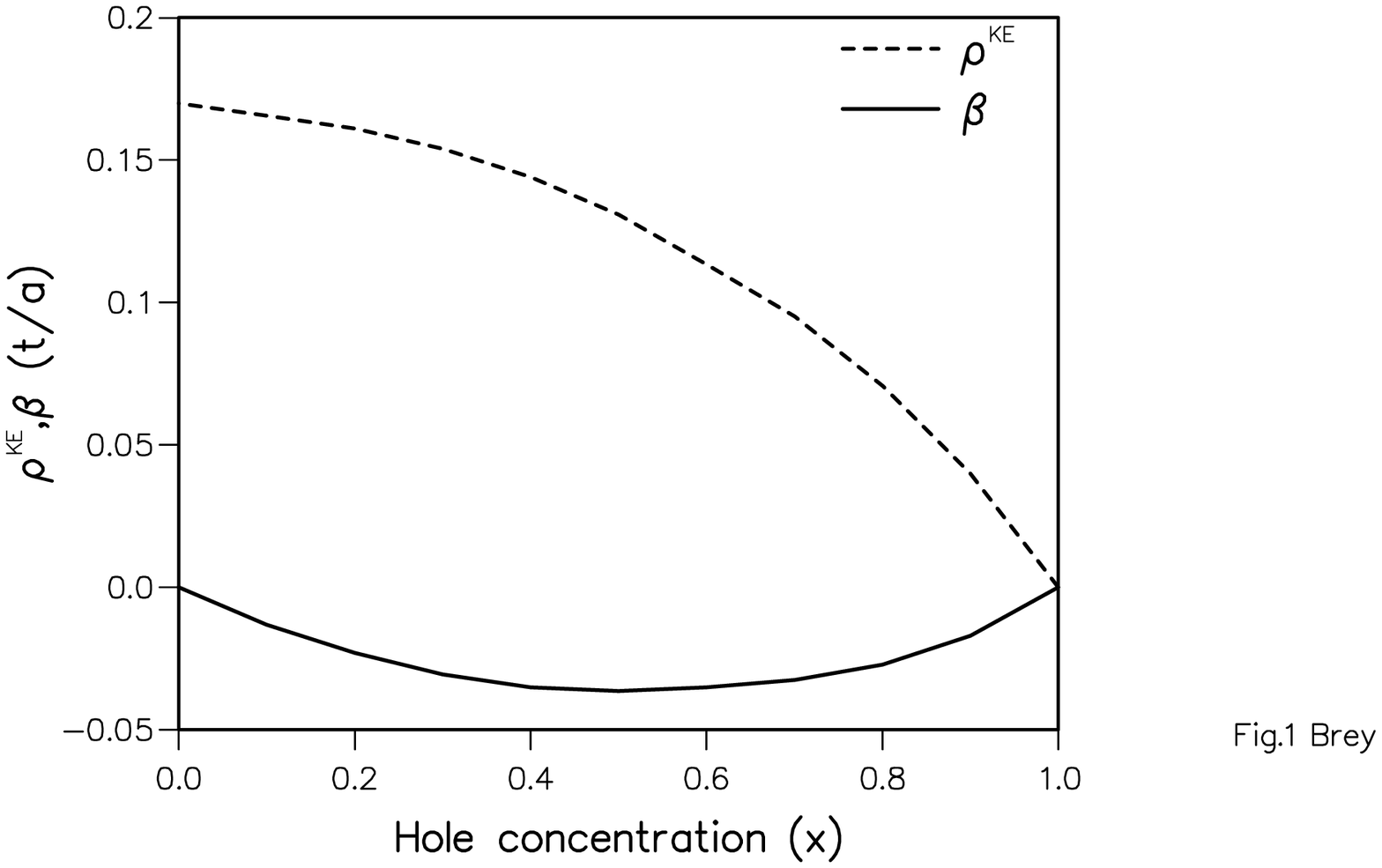,width=11.0cm}
\caption{
Variation of a) $\rho ^{KE}$ and b) $\beta$ as a function of $x$, as
obtained from Hamiltonian Eq.1.
Note that $x$ is measured with respect half filling.}
\end{figure}
\vspace{-7cm}
\begin{figure}
\epsfig{figure=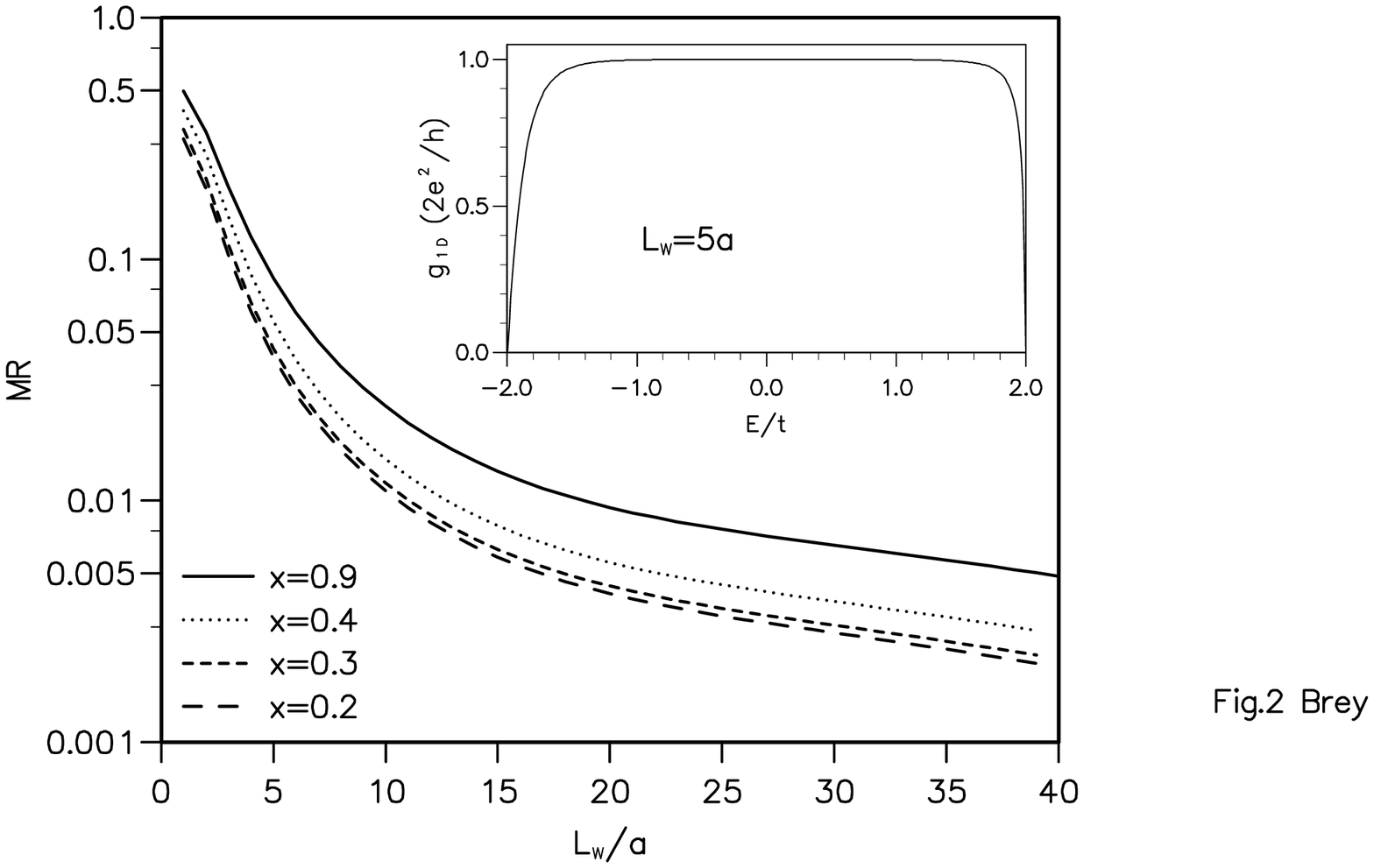,width=11.0cm}
\caption{
Variation of $MR$ as a function of $L_W$ for 
different hole concentrations.
The  inset  shows  the one-dimensional conductance as a function of the
energy,  for a DW with $L_W$=5.}
\end{figure}


\begin{references}
%\bibitem{prim}R.M. Kusters, J. Singleton, D.A. Keen, R. McGreevy and W. Hayes, 
%Physica (Amsterdam) {\bf 155B}, 
%362 (1989); K. Chahara, T. Ohno, M. Kasai and Y. Kozono, 
%Appl. Phys. Lett. {\bf 63}, 1990 (1993);
%R. von Helmolt, J. Wecker, B. Holzapfel, L. Schulzt and K.
%Samwer,  Phys. Rev. Lett. {\bf 71}, 2331 (1993);
%S. Jin, T.H. Tiefel, M. McCormack, R.A. Fastnacht, R. Ramesh and J.H. Chen,
%Science {\bf 264}, 413 (1994).
\bibitem{prim}R.M. Kusters {\it et al.}
Physica (Amsterdam) {\bf 155B}, 362 (1989); 
K. Chahara {\it et al.} Appl. Phys. Lett. {\bf 63}, 1990 (1993);
R. von Helmolt {\it et al}, Phys. Rev. Lett. {\bf 71}, 2331 (1993);
S. Jin {\it et al}, Science {\bf 264}, 413 (1994).


\bibitem{rvw}See, for example, A.P. Ramirez, J. Phys.: Condens. Matter {\bf 9},
8171(1997);
J.M.D. Coey, M. Viret and S. von Molnar, Adv. in Phys. {\bf 48}, 167, (1999).
\bibitem{zener}C. Zener, Phys. Rev. {\bf 82}, 403 (1951).
%\bibitem{anderson}P.W. Anderson and H. Hasegawa, 
%Phys. Rev {\bf 100}, 675 (1955).
\bibitem{anderson}P.W. Anderson {\it et al.}
Phys. Rev {\bf 100}, 675 (1955).
\bibitem{degennes}P.G. deGennes, Phys. Rev. {\bf 118}, 141 (1960).
\bibitem{note1}A full understanding of the metal-insulator transition 
requires to add other terms to the DE model, in particular  electron-phonon
coupling: 
%A.J.Millis, P.B.Littlewood and B.I. Shraiman, Phys.Rev.Lett.
A.J.Millis {\it et al.} Phys.Rev.Lett.  {\bf 74}, 5144 (1995).
%\bibitem{grain1}
%H.Y.Hwang, S.W.Cheong, N.P.Ong and B.Batlogg, Phys.Rev.Lett, {\bf 77}, 24041 (1996);
%A.Gupta, G.Q.Gong, G.Xiao, P.R.Duncombe, P.Lecoeur,
%P.Trouilloud, Y.Y.Wang, V.P.Dravid, and J.Z.Sun, 
%Phys.Rev.B {\bf 54}, R15629 (1996); 
%G.J.Snyder, R.Hiskes, S.DiCarolis, M.R.Beasley, and T.H.Geballe,
%Phys.Rev.B {\bf 53}, 14434 (1996);
%Ll.Balcells, J.Fontcuberta, B.Mart\'{\i}nez and X.Obradors Phys.Rev.B in press.
%HIGH-FIELD MAGNETORESISTENCE AT INTERFACES IN MANGANESE....
%\bibitem{films1}N.D.Mathur, G.Burnell, S.P.Isaac, T.J.Jackson, B.-S Teo, J.L.MacManus-Driscoll, L.F.Cohen, J.E.Evetts and M.G.Blamire, Nature {\bf 387}, 266 (1997);
%K.Steenbeck, T.Eick, K.Kirsh, K.O'Donell and E.Stenbeiss, 
%Appl.Phys.Lett. {\bf 71}, 968 (1997).
\bibitem{grain1}
H.Y.Hwang {\it et al.} Phys.Rev.Lett, {\bf 77}, 24041 (1996);
A.Gupta {\it et al.}
Phys.Rev.B {\bf 54}, R15629 (1996); 
G.J.Snyder {\it et al.}
Phys.Rev.B {\bf 53}, 14434 (1996);
Ll.Balcells {\it et al.} Phys.Rev.B {\bf 58}, R14697 (1998).
%\bibitem{films1}N.D.Mathur, G.Burnell, S.P.Isaac, T.J.Jackson, B.-S Teo, J.L.MacManus-Driscoll, L.F.Cohen, J.E.Evetts and M.G.Blamire, Nature {\bf 387}, 266 (1997);
%K.Steenbeck, T.Eick, K.Kirsh, K.O'Donell and E.Stenbeiss, 
%Appl.Phys.Lett. {\bf 71}, 968 (1997).
\bibitem{films1}N.D.Mathur {\it et al.} Nature {\bf 387}, 266 (1997);
K.Steenbeck {\it et al.}
Appl.Phys.Lett. {\bf 71}, 968 (1997).

%\bibitem{tri1}
%J.Z.Sun, W.L.Gallagher, P.R.Duncombe, L.Krusin-Elbaurn, R.A.Altman, A.Gupta
%Y.Lu, G.Q.Gong, and G.Xiao, Appl.Phys.Lett. {\bf 69}, 3266 (1996);
%Y.Lu,X.W.Li,G.Q.Gong, G.Xiao, A.Gupta, P,Lecoeur, J.Z.Sun, Y.Y.Wangm ans
%V.P.Dravid, Phys.Rev.B {\bf 54}, R8357 (1996).

\bibitem{tri1}
J.Z.Sun {\it et al.}
Appl.Phys.Lett. {\bf 69}, 3266 (1996);
Y.Lu {\it et al.}
Phys.Rev.B {\bf 54}, R8357 (1996).

\bibitem{calderon1}M.J.Calder\'on, L.Brey and F.Guinea, cond-mat/9811337.
%\bibitem{iti}
%K.Hong and N.Giordano, Phys.Rev.B {\bf 51} 9855 (1995);
%J.F.Gregg, W.Allen, K.Ounadjela, M.Viret, M.Hern, S.M.Thompson and J.M.D.Coey
%Phys.Rev.Lett. {\bf 77}, 1580 (1996);
%Y.Otani, S.G.Kim, K.Fukamichi, O.Kitakami and Y.Shimada, 
%IEEE Trans.Magn. {\bf 34}, 1096 (1998);
%A.D.Kent, U.Ruediger, J.Yu, L.Thomas and S.S.P.Parkin J.Appl.Phys. in press.
\bibitem{iti}
K.Hong {\it et al.}, Phys.Rev.B {\bf 51} 9855 (1995);
J.F.Gregg {\it et al.}
Phys.Rev.Lett. {\bf 77}, 1580 (1996);
Y.Otani {\it et al.}
IEEE Trans.Magn. {\bf 34}, 1096 (1998);
A.D.Kent {\it et al.} J.Appl.Phys. in press.


%\bibitem{nico}N.Garc\'{\i}a, M.Mu\~noz and Y.-W.Zhao,
%Phys.Rev.Lett. {\bf 82}, 2923 (1999).
\bibitem{nico}N.Garc\'{\i}a {\it et al.}
Phys.Rev.Lett. {\bf 82}, 2923 (1999).
%\bibitem{mathur}
%N.D.Mathur, P.B.Littlewood, N.K.Tood, S.P.Isaac, B.-S.teo, D.-J.Kang,
%W.J.Tarte, Z.H.Barber, J.E.Evetts and M.G.Blamire, J.Appl.Phys. in press.
\bibitem{mathur}
N.D.Mathur {\it et al.}
submitted to rJ.Appl.Phys.
%\bibitem{itith}
%G.Tatara and H.Fukuyama, Phys.Rev.Lett. {\bf 67}, 3773 (1997);
%J. van Hoof, K.M.Schep, A.Brataas, G.E.W.Bauer and P.J.Kelly
%Phys.Rev.B {\bf 59} 138 (1999).
\bibitem{itith}
G.Tatara {\it et al.}, Phys.Rev.Lett. {\bf 67}, 3773 (1997);
J. van Hoof {\it et al.}
Phys.Rev.B {\bf 59} 138 (1999).
%\bibitem{yamanaka}M.Yamanaka and N.Nagaosa J.Phys.Soc.Jpn. {\bf 65}, 
%3088 (1996).
\bibitem{yamanaka}M.Yamanaka i{\it et al.} J.Phys.Soc.Jpn. {\bf 65}, 
3088 (1996).
%\bibitem{muller}E.M\"uller-Hartmann and E. Dagotto, 
%Phys. Rev. B {\bf 54}, R6819 (1996).
\bibitem{muller}E.M\"uller-Hartmann {\it et al.}
Phys.Rev.B {\bf 54}, R6819 (1996).
\bibitem{oguri}A.Oguri, Phys.Rev.B {\bf 56}, 13422 (1997).
\bibitem{calderon2}M.J.Calder\'on {\it et al.}
Phys.Rev.B {\bf 59},4170 (1999).
\bibitem{hirota}K.Hirota {\it el al.} J.Phys.Soc.Jpn. {\bf 65}, 3736 (1996).
\bibitem{hwang}H.Y.Hwang {\it et al.} Phys.Rev.Lett. {\bf 80}, 1316 (1998).
\bibitem{perring}T.G.Perring {\it et al.}Phys.Rev.Lett. {\bf 77}, 711 (1996).
\bibitem{sanchez}R.S\'anchez {\it et al.} Material Science Forum, 
{\bf 235}, 831 (1997).
\bibitem{bruno}P.Bruno cond-mat/9904370.
\bibitem{ott}F.Ott {\it et al.} Phys.Rev.B {\bf 58}, 4656 (1998).



\end{references}
\end{document}